\documentclass[preprint,5p,12pt,twocolumn]{article}
\usepackage{flushend}
\usepackage{amssymb}
\usepackage{subfig}
\usepackage{algorithm,algorithmic}
\usepackage{graphicx}
\usepackage{caption}
\usepackage{setspace}
\usepackage{xspace}
\usepackage{multirow}
\usepackage{rotating}
 \usepackage{url}
\usepackage{gensymb}
\usepackage [ table ]{ xcolor }
\usepackage{natbib}
\usepackage[top=2cm, bottom=2cm, left=1.5cm, right=1.5cm]{geometry}
\usepackage[affil-it]{authblk}

\begin{document}
\makeatletter
\newcommand*{\rom}[1]{\expandafter\@slowromancap\romannumeral #1@}
\makeatother

\title{Object categorization in finer levels requires higher spatial frequencies, and therefore takes longer}

%% use optional labels to link authors explicitly to addresses:
%% \author[label1,label2]{}
%% \address[label1]{}
%% \address[label2]{}
\author{Matin N. Ashtiani$ ^{1} $}
\author{Saeed Reza Kheradpisheh$ ^{1,2} $ }
\author{Timoth\'ee Masquelier$ ^{3} $}
\author{Mohammad Ganjtabesh$ ^{1,2,} $\thanks{Corresponding author.\\ Email addresses:\\ matin.ashtiani@ut.ac.ir (MNA),\\ kheradpisheh@ut.ac.ir (SRK),\\ timothee.masquelier@cnrs.fr (TM),\\ mgtabesh@ut.ac.ir (MG).}}

\affil{\footnotesize $ ^{1} $ Department of Computer Science, School of Mathematics, Statistics, and Computer Science, University of Tehran, Tehran, Iran}
\affil{\footnotesize $ ^{2} $ School of Biological Sciences, Institute for Research in Fundamental Sciences (IPM), Tehran, Iran}
\affil{\footnotesize $ ^{3} $ CERCO UMR 5549, CNRS – Universit\'e de Toulouse 3, F-31300, France} 
\date{}
%\cortext[cor1]{Corresponding author}
%\address[UT]{Department of Computer Science, School of Mathematics, Statistics, and Computer Science, University of Tehran, Tehran, Iran}
%\address[Inserm]{INSERM, U968, Paris, F-75012, France.}
%\address[Sorbonne]{Sorbonne Universit\'es, UPMC Univ Paris 06, UMR-S 968, Institut de la Vision, Paris, F-75012, France.}
%\address[CNRS]{CNRS, UMR-7210, Paris, F-75012, France}
\maketitle

\begin{abstract}

The human visual system contains a hierarchical sequence of modules that take part in visual perception at different levels of
abstraction, i.e., superordinate, basic, and subordinate levels. One important question is to identify the
``entry" level at which the visual representation is commenced in the process of object
recognition. For a long time, it was believed that the basic level had advantage
over two others; a claim that has been challenged recently.
Here we used a series of psychophysics experiments, based on a rapid presentation paradigm, as well as two computational models, with bandpass filtered images to study the processing order of the categorization levels.
In these experiments, we investigated the type of visual information required for categorizing objects in each level
by varying the spatial frequency bands of the input image. The results of our psychophysics experiments and computational models are consistent. They indicate that the different spatial frequency information had different effects on object categorization in each level.
In the absence of high frequency information, subordinate and basic level categorization are performed inaccurately, while superordinate level is performed well. This means that, low frequency information is sufficient for superordinate level, but not for the basic and subordinate levels. These finer levels require high frequency information, which appears to take longer to be processed, leading to longer reaction times. Finally, to avoid the ceiling effect, we evaluated the robustness of the results by adding different amounts of noise to the input images and repeating the experiments. As expected, the categorization accuracy decreased and the reaction time increased significantly, but the trends were the same.This shows that our results are not due to a ceiling effect.

\textbf{Keywords:} Spatial frequencies, Object categorization, Categorization levels, Psychophysics, Rapid object presentation  %All article types: you may provide up to 8 keywords; at least 5 are mandatory.
\end{abstract}

\section{Introduction}
An object can be categorized in different levels of abstraction, including the superordinate (e.g., animal), basic (e.g., bird), and subordinate (e.g., duck) levels. The processing order of these levels is yet being debated. There are several studies suggesting that categorization in the basic level is completed prior to the superordinate level~\citep{tanaka1991object,rogers2007object,collin2005subordinate,dehaqani2016temporal}. On the other side, the advantage of the basic level has been challenged by showing faster
visual processing for the superordinate level in rapid-presentation experiments. Using two-forced-choice behavioral experiments with long (i.e., 500ms) and short (i.e., 50ms) presentation times, \cite{bowers2008detecting} showed that superordinate level categorization (object/texture images) is completed before the basic level (e.g. dog/bus). Also, \cite{mace2009time} found that rapidly presented (26ms) natural images were faster to be categorized at the superordinate level than the basic level. Using a forced-choice saccadic task, \cite{wu2014120} found that humans can accurately perform superordinate level categorization at 120ms, while the accuracy of basic level categorization is around chance-level. Although  \cite{mack2015dynamics} challenged the rapid presentation paradigm for studying the processing order of categorization levels, studies done by \cite{poncet2014stimulus} and \cite{vanmarcke2016time} showed that the advantage  of superordinate level is not affected by the stimulus duration (25ms to 500ms) and diversity. Also, \cite{prass2013ultra} showed that the background context and animacy have no effect on the superordinate level advantage. 

There is evidence indicating that the visual system processes visual input in order from low to high spatial frequencies~\citep{schyns1994blobs, mace2005entry,kauffmann2015neural}, or from general shape to fine details. However, it is still disputed  how brain processes different spatial frequencies~\citep{kauffmann2014neural}. \cite{bar2006top} suggests that low spatial frequencies (LSFs) are analyzed quickly and provide an initial and general guess about the object, which then facilitates the object categorization. Indeed, it is suggested that the LSFs are rapidly conveyed by the magnocellular pathways into the high cortical areas (e.g., orbitofrontal cortex). There, a coarse object representation is formed and then back-projected to the inferiortemporal cortex to refine the subsequent processing of high spatial frequencies (HSFs) conveyed by the parvocellular pathways through the ventral visual cortex~\citep{bar2006top,kauffmann2015neural}.

Therefore, studying the impact of spatial frequencies on humans' accuracy and reaction time (RT) in object categorization tasks at different levels can help to unravel the entry categorization level challenge. Using speeded category verification tasks, \cite{collin2005subordinate} showed that basic level categorization is completed earlier than superordinate and subordinate levels. The main critic to these types of experiments is the use of semantic labels, that involves semantic processing of the brain~\citep{wu2014120}. Binding the object visual representation and its name takes time which may be different for each categorization level~\citep{mace2009time}. Here, we used a rapid presentation paradigm with frequency-filtered images to study the processing order of the categorization levels. Indeed, subjects were asked to determine the category of the object image presented for a duration of 12.5ms in one of the superordinate, basic, and subordinate levels, when the image was intact or bandpass filtered into one of the LSF, HSF, or intermediate spatial frequency (ISF) bands. For each categorization level, we performed several psychophysics tasks in which images of different object categories were used. This way, we could check whether the results are independent from target object categories.

The results of our psychophysics experiments indicated that the superordinate level categorization mainly relies on LSFs, while the basic and subordinate levels require higher spatial frequencies. Indeed, for superordinate level, the human categorization accuracy peaks at LSF band and drops with the ISF and HSF bands. On the contrary, for the basic and subordinate levels, the accuracy increases by increasing the spatial frequency (with greater slope in subordinate level). However, the RT always increases with the spatial frequency, whatever the categorization level, which is compatible with the processing order of spatial frequencies (from low to high). Also, RTs decrease with categorization level, whatever the frequency band. These findings are in support of the temporal advantage of superordinate to the basic level, as well as, basic to the subordinate level.

We also evaluated two object categorization models on the same set of experiments in different categorization levels using images in different frequency bands. This helps to investigate whether the results of our psychophysics experiments are due to the specific processing mechanisms of the visual system or they are forced by the information content in each frequency band. Interestingly, the categorization accuracies of both models strongly correlate with human accuracies in all categorization levels and frequency bands. For the LSF band, both models could accurately perform superordinate categorization, and failed on basic and subordinate levels. For intermediate and HSF bands, the models' accuracy drops for the superordinate level, but increases for the basic and subordinate levels (with greater slope in subordinate level). This suggests that, from a computational point of view, the LSF band carries sufficient visual information to perform superordinate level categorization, while for the basic and subordinate levels, higher frequencies are required. Thus, since lower spatial frequencies are processed earlier than higher ones, the superordinate level appears to be the entry categorization level, and subsequently, the basic and subordinate levels are completed.

Further, we added different amounts of phase noise to the images and repeated all the psychophysics and computational experiments, to check for any possible ceiling effect. By increasing the noise level, the accuracies (resp. RTs) in all categorization levels and frequency bands are decreased (resp. increased). However, at each noise level, the trend of the accuracies and RTs is the same as the noise-less experiments. Again, this confirms that our results are caused by the visual information at each frequency band.

\section{Materials and methods}
\subsection{Dataset}
\label{sec:dataset}
\begin{figure*}[!htb]
\begin{center}
\includegraphics[scale = 0.75]{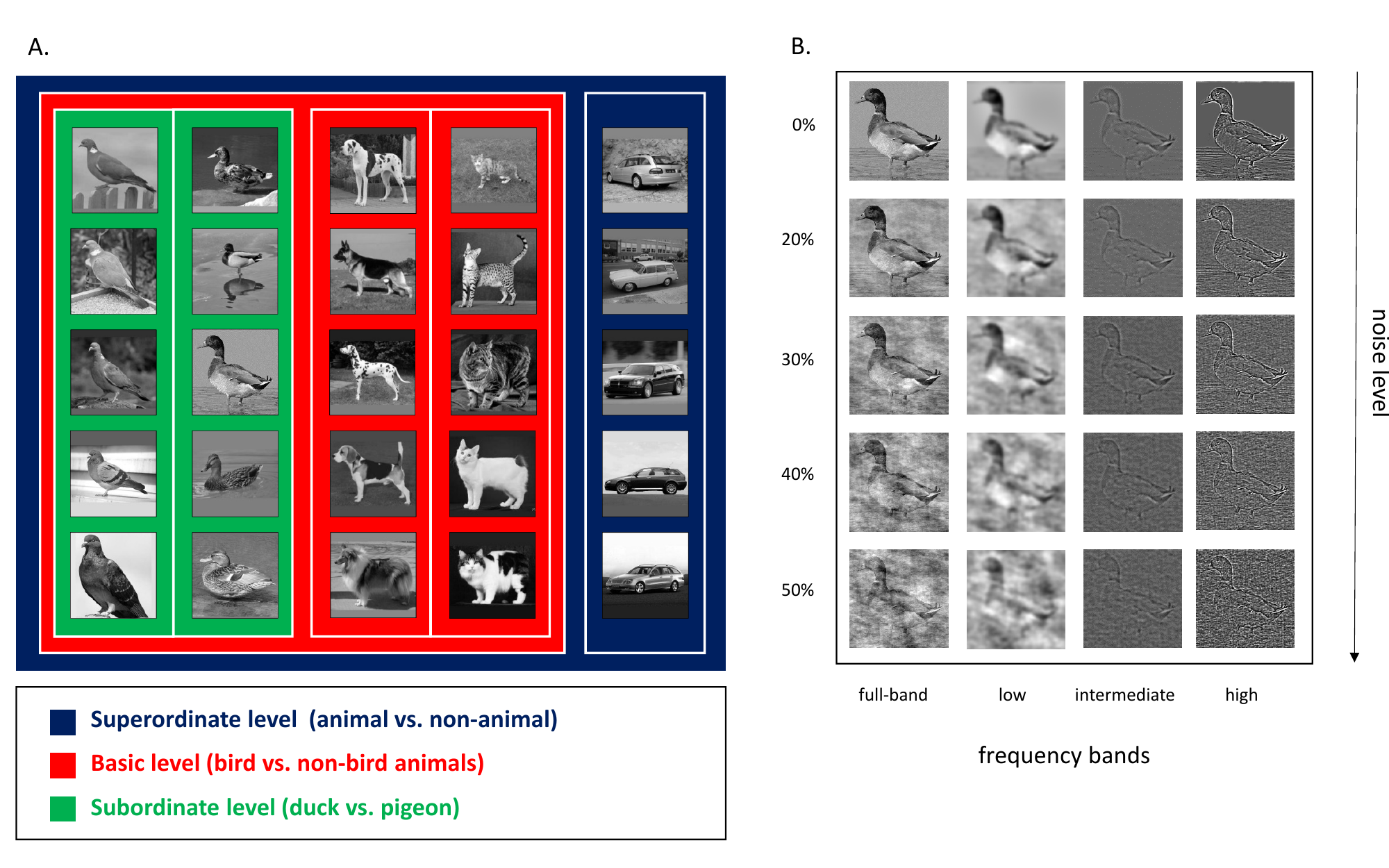}
\end{center}
\caption{
A. Examples of image stimuli and the hierarchy of three categorization levels. Levels are specified by different colors, i.e., superordinate in blue, basic in red and subordinate in green borders. B. A sample image bandpass filtered into different spatial frequency bands and contaminated with different amounts of phase noise. Columns specify different frequency bands: full-band (unfiltered image), LSF, ISF, and HSF bands respectively from left to right. Rows correspond to the noise levels: 0\% (without noise), 20\%, 30\%, 40\%, and 50\% phase noise respectively from top to bottom.
%The obtained images results of applying frequency filter using Furier transform (low, intermediate and high spatial frequencies) and unfiltered images in each  and phase noise at different conditions (0\%, 20\%, 30\%, 40\% and 50\% phase noise); 
}\label{fig:dataset}
\end{figure*}

We used images of five object categories, including ducks, pigeons, cats, dogs, and cars (200 images per category). Most of the images were picked from the Imagenet dataset~\citep{ILSVRC15}, and others were gathered from the web. Each image contains the side view of a different instance of one of the object categories. For each categorization level, a different combination of categories has been used. For the superordinate level (animal vs. non-animal), there was four different sets of image in two categories: one of the four animals and the car category. Also, for the basic level (bird vs. non-bird animals) there was also four sets, each of which containing one bird (duck or pigeon) and one non-bird animal (cat or dog). In the subordinate level, only duck and pigeon categories were employed. Figure~\ref{fig:dataset} shows some examples from each category as well as the hierarchy of the categorization levels. Note that the images were grayscaled and cropped to have 300$\times$300 pixels.

All the images have also been filtered into LSF, ISF, and HSF bands. To produce the filtered images, first, each original image was Fourier transformed into the frequency domain, and then, multiplied by a 2D frequency filter to keep the desired frequencies, and finally, backed to the spatial domain using inverse Fourier transformation. Here, we used 2D Gaussian low-pass function to construct the desired frequency filters. The general form of a Gaussian low-pass, $H_{LP}$, and Gaussian high-pass, $H_{HP}$, filters are as follows:
\begin{equation}
    H_{LP}(u,v)=exp\left(-\frac{(u/M)^2+(v/N)^2}{2.F_{l}^2}\right),
\end{equation}
and 
\begin{equation}
    H_{HP}(u,v)=1-H_{LP}(u,v),
\end{equation}
where $u$ and $v$ are the frequency coordinates, $M$ and $N$ are correspondingly the maximum frequency component at each frequency dimension, and $0 \leq F_{l} \leq 1$ is the frequency cut-off rate. Using two Gaussian filters, say $H_{LP1}$ and $H_{LP2}$ with the corresponding cut-off rates of $F_{l1}$ and $F_{l2}$ ($F_{l1}<F_{l2}$), the band-pass filter, $H_{BP}$, is calculated as:
\begin{equation}
    H_{BP}(u,v)=H_{LP2}(u,v)-H_{LP1}(u,v),
    \label{eq:3}
\end{equation}

To prepare the LSF-, ISF-, and HSF-filtered images, we used respectively $H_{LP1}$, $H_{BP}$ and $H_{HP2}$ ($=1-H_{LP2}$) filters with cut-off frequency rates of $F_{l1}=0.25$ and $F_{l2}=0.60$.
 
We also prepared a noisy version of the original and frequency-filtered images. We added phase noise in different levels (20\%, 30\%, 40\%, and 50\%) to each image. Unlike other noise generating methods (e.g., simple white noise), the phase noise produces a noise signal that is proportional to the energy of image at each spatial frequency level. Indeed, it consists of frequency components of the image that have been displaced, therefore, the phase noise will have exactly the same energy distribution as the image itself. We used a noise addition mechanism analogous to \cite{ales2012objective} study. 
%To produce the noise signal at each level, the phase of each frequency component is displaced by a random value in the corresponding range.
%For each image, we interpolated between the starting unscrambled image and the final image that had 100\% randomized phases. A phase-scrambling parameter used to systematically vary the visibility of the image. We varying this parameter from 0\% to 100\% in 10\% increments, using a linear phase interpolation technique and finally only the 20\%, 30\%, 40\%, and 50\% phase scrambled images took into account for experiments.

%For instance, for the 20\% level, the phase of each component will  change, at most, about 20\%. 

\subsection{Psychophysics experiments}
\label{sec:methods_psycghophsyics}
We performed 12 rapid two-forced-choice object categorization experiments containing three categorization levels (superordinate, basic, and subordinate), two image types (i.e., original and frequency-filtered images), and two noise conditions (i.e., images with and without noise).  Each trial started with a fixation point presented on a uniform gray background for 500ms. Then, an stimulus image was shown for 12.5ms (one frame on a 80Hz monitor) followed by a uniform gray screen, presented for another 12.5ms, as an inter-stimulus interval (ISI). Immediately afterwards, a 1/f noise mask was shown for 150ms. Finally, subjects should report the category of the stimulus image, by pressing the corresponding key on a keyboard. Each experiment session started with a training phase in which subjects learned to do the categorization task in the desired level, followed by a recording phase in which we recorded the subjects' RT and performance. The training phase of each session contained 20 images (10 images per category) that are randomly selected. When subjects reported their decision, a feedback was shown to them indicating whether they responded correctly or not. The recording phase contained 240 trials (120 images per category) without any feedback. Images used in the training phase are not shown in the recording phase.

Subjects were seated on a comfortable chair in a dark room and were instructed to respond as fast and accurate as possible. Stimuli were presented using Matlab Psychophysics Toolbox~\citep{brainard1997psychophysics} in a $21"$ CRT monitor with a resolution of 800${\times}$600 pixels, frame rate 80Hz, and viewing distance of 60cm. Therefore, each stimulus covered $11^{\circ}\times11^{\circ}$ of visual angle. All subjects voluntarily participated to the experiments and gave their written consent prior to participation. All experimental protocols were approved by the ethical committee of the University of Tehran. All experiments were carried out in accordance with the guidelines of the declaration of Helsinki and the ethical committee of the University of Tehran.

%It should be noted that the stimuli were presented in a dark room, on a CRT monitor (800${\times}$600 pixels, refresh rate 80Hz) and subjects were seated around 60 centimeters from the monitor (11$^{\circ}$ ${\times}$11$^{\circ}$ visual angle).

For the superordinate level experiments, there were four animal/non-animal tasks (one of the four animals vs. car). Also, there were four bird/non-bird tasks (duck or pigeon v.s cat or dog) for the basic level, and one bird categorization task (duck vs. pigeon) for the subordinate level experiments. Regarding the image type and noise condition, we classify all the experiments in four groups:
\begin{itemize}
    \item {\textbf{Original images (i.e., full-band):}}\\ In these experiments, the original images were used as stimuli. 40 subjects participated in the superordinate level experiment (10 for each task), 40 subjects performed the basic level experiment (10 for each task), and 20 subjects did the subordinate level experiment. Images were randomly shuffled and shown in different trials.
    
    \item {\textbf{Frequency-filtered images:}}\\ For these experiments, we used filtered images in LSF, ISF, and HSF bands (see Section~\ref{sec:dataset}).  The number of subjects who participated in superordinate, basic, and subordinate level experiments were the same as in the ``Original images" case. For each frequency band, 40 images per category were used in the recording phase of each experiment session (2 [category] $\times$ 3 [frequency bands] $\times$ 40 [images] = 240 images). For the training phase, we also used frequency-filtered images. Notably, images were presented in a random order.
    
    \item {\textbf{Noisy images:}}\\ In these experiments, we used the noisy version of the original images (see Section~\ref{sec:dataset}) with four different noise levels (20, 30, 40, and 50 percent). Here again, we had 40 subjects  for each of the superordinate and basic level experiments and 20 subjects for the subordinate level. We used 30 images for each noise level, and thus, in total 120 images per category were presented in each task (2 [category] $\times$ 4 [noise levels] $\times$ 30 [image] = 240 images). Similar to the previous case, the order of images was random.
        
    \item {\textbf{Frequency-filtered noisy images:}}\\ Frequency-filtered images contaminated with noise (Section~\ref{sec:dataset}) were used in this series of experiments. There again, the number of subjects in superordinate, basic, and subordinate level experiments were the same as in the ``Original images" case. For each frequency band and noise level, 10 images per category were used in the recording phase of each experiment session (2 [category] $\times$ 3 [frequency bands] $\times$ 4 [noise levels] $\times$ 10 image= 240 images). For the training phase, we also used frequency-filtered images in different noise levels. Images were presented in random order.
\end{itemize}

\subsection{Computational models}
\label{sec:methods_models}
To investigate the information content in each frequency band for the different categorization levels, we used two computational models. Each model was evaluated on similar categorization tasks as performed in our psychophysics experiments. Indeed, we performed the superordinate, basic, and subordinate level categorization tasks with the full-band, frequency-filtered, noisy, and frequency-filtered noisy images, separately. Thus we had 12 experiments (3 [categorization levels]$\times$2 [image types]$\times$2 [noise condition]). Again, for the superordinate level experiments, we performed four animal/non-animal tasks. Also, for the basic and subordinate level experiments, we performed four bird/non-bird animal and one bird categorization tasks, respectively. For each categorization task, we used 400 training images (200 per category) for extracting features and training the classifier, and 400 test images to evaluate the model. In each task, both the training and testing images belonged to the same group. For instance, if the experiment was performed on LSF-filtered images, then all the training and testing samples were filtered in LSF band. However, the frequency filtering mechanism depends on the structure of each model (see below). It should be mentioned that for both models, we used grayscaled images that were rescaled to have 140 pixels in height.

\begin{figure*}[!htb]
\begin{center}
\includegraphics[scale = 0.5]{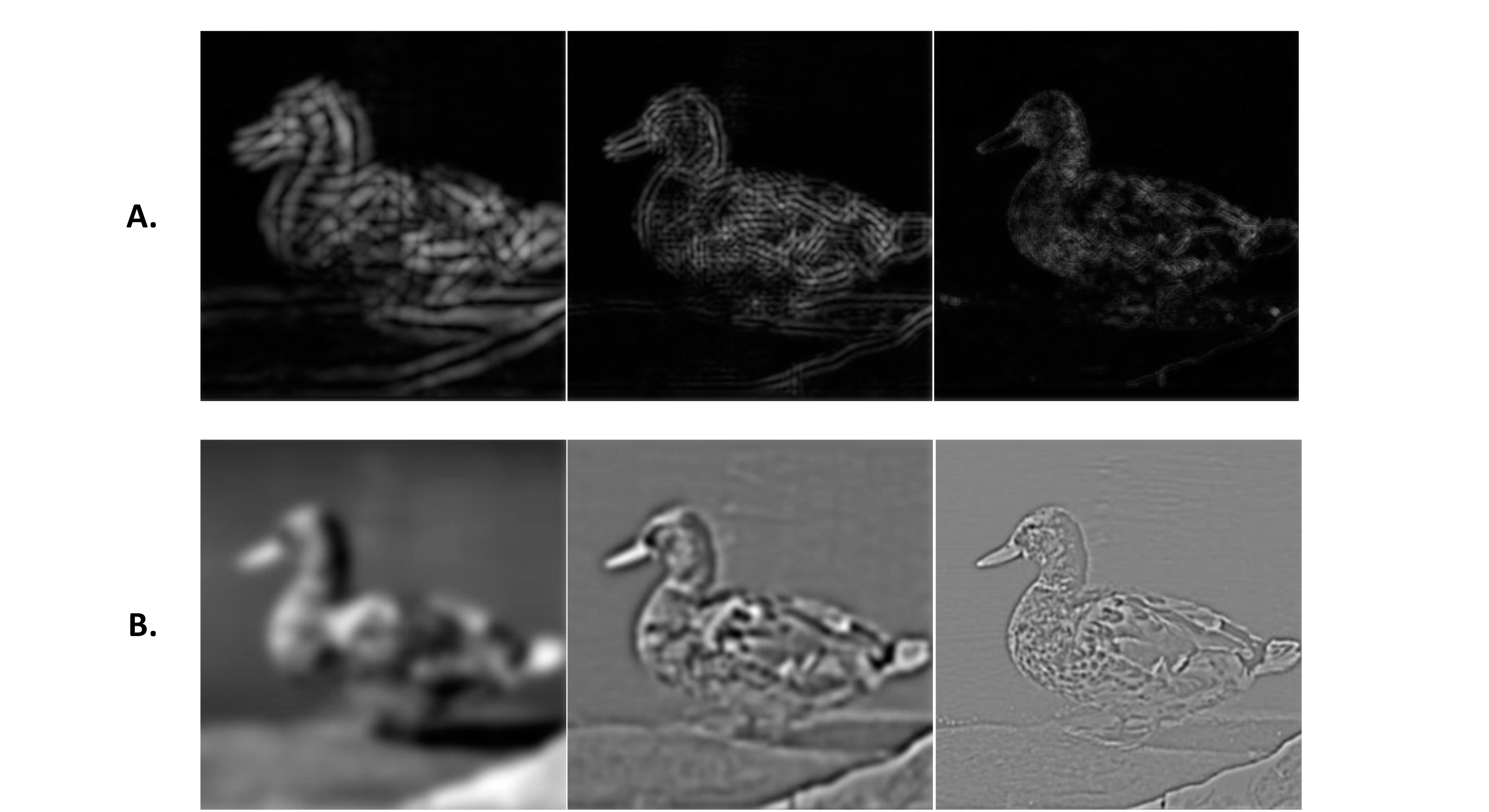}
\end{center}
\caption{
 A sample image filtered into different spatial frequency bands using Gabor filters and Gaussian bandpass filters. Image filtered into the LSF, ISF, and HSF bands are respectively shown from left to right. A. Image of each frequency band is the accumulation of the outputs of Gabor filters in four orientations (0, 45, 90, and 135 degrees). B. Gaussian bandpass filtered images using $H_{LP1}$, $H_{BP}$ and $H_{HP2}$ filters (see Section~\ref{fig:dataset}).}\label{fig:input}
\end{figure*}

\subsubsection{Model \rom{1}}
The first model is largely inspired by the HMAX model \citep{serre2007robust}, which is widely used in the visual neuroscience studies as a model of object recognition. In this model, the input image is first filtered in the S1 layer, by applying various Gabor filters with different spatial frequencies and orientations. Then, in the C1 layer, a local max operation is performed over the output of S1 layer. Afterwards, random patches are extracted from the output of C1 layer over the training images. These patches are considered as object representative visual features corresponding to the object categories. For each test image, all extracted patches are convolved with the output of C1 layer in different positions (S2 layer) and the maximum convolution values, corresponding to the extracted visual features, provide the object representation in C2 layer. Finally, a classifier detects the category of the input image based on its C2 representation.

The standard HMAX model uses Gabor filters with 16 different spatial frequencies and four orientations in the S1 layer (i.e., 0, 45, 90, and 135 degrees), which are compatible with the recordings in area V1; see~\citep{serre2007robust} for more details. In the C1 layer of HMAX, the local max operation is performed over a neighbourhood of two adjacent frequencies, i.e., the C1 layer compresses every two consecutive S1 maps into one C1 map. We used the first two C1 maps as high, the next four as intermediate, and the last two as LSF bands. Hence, for each frequency band, the original images are fed into the model and the corresponding Gabor filters are applied on them in the S1 layer. Then, the C1 maps are computed by performing a local max operation over the output of two consecutive S1 maps. For instance, in the HSF band, we used Gabor filters in the first four spatial frequencies, while the other frequencies are totally neglected.

In the S2 layer, we picked random patches from each C1 map, where the size of these patches varied from 4$\times$4 to 24$\times$24 with a step of 2. Then, for each patch size, 1000 random patches are extracted from different training images. Therefore, for instance, the S2 layer crops 22000 random patches (2 [C1 maps] $\times$ 11 [patch sizes] $\times$ [1000 patches per size]) for the HSF band. In C2 layer, we perform a global max operation over each S2 map and put them together as the representative feature vector. Finally, we employed a 1-nearest neighbor (1-NN) classifier to categorize the input test image based on the label of the closest training sample in C2 feature space. 

Therefore, for the experiments with the full-band images (i.e., original images) we used all the eight C1 feature maps. But, for the LSF, ISF, and HSF cases, we only used the corresponding C1 maps. In the noisy image experiments, first we added the noise to the input image, and then fed it to the model.
%For instance, in the low frequency-filtered noisy image case, we first add the noise to the image, and then, we only use the last two C1 output maps.

\subsubsection{Model \rom{2}}

The frequency filtering mechanism applied in model~\rom{1} (using Gabor filters with different spatial frequencies) is different from what was used in the psychophysics experiments. Therefore, using model~\rom{2}, which replaces S1 and C1 layers with directly frequency-filtered images, we could verify that the results are independent from the frequency filtering mechanism.

In this model, we discarded the first two layers of the standard HMAX. Thus, for frequency-filtered image experiments, we used the same procedure as explained in Section~\ref{sec:dataset} to filter images into the desired frequency bands. For the full-band, we extracted 11000 patches (11 [patch sizes]$\times$1000 [patches per size]) directly cropped from the original images. For the Frequency-filtered (noisy) images, the patches are extracted from the filtered (noisy) images.  Afterwards, to construct the feature vectors, we convolved these patches with the input images and performed a global max operation. At the end, a similar classifier (1-nearest neighbor) is used for deciding about the category of the input image.

%\subsection*{Analysis}
%Reaction times (RT) and the proportion of correct responses ($\%$-correct) were calculated. Statistical analysis was performed by using IBM SPSS statistics 22 (version 22.0.0.0). To analyze the obtained results repeated measures Analysis of Variance (ANOVA) and post hoc pairwise comparisons (Tukey test) were conducted. Error bars of graphs represent the standard error of the mean(SEM).
%
%The percentages of correct responses as well as the reaction times were analyzed with a 3 (level: superordinate, basic, subordinate)$\times$4 (frequency: low, intermediate, high, original image)$\times$9 (task: 4 superordinate, 4 basic and 1 subordinate tasks) Analysis of Variance (ANOVA).

\section{Results}
In this section, we present the results of the psychophysical and computational experiments. Section~\ref{sec:results_psycghophsyics} provides the accuracy and reaction time of human subjects performing the superordinate, basic, and subordinate level psychophysics experiments with the original and frequency-filtered images. Then, in Section~\ref{sec:results_models}, the recognition accuracy of both computational models over the same experiments is presented. Finally, the robustness of the results to different amounts of noise is examined in Section~\ref{sec:results_noise}.

%It should be noted that these accuracies are the results of ten times execution of the models and the average accuracies are reported.

\subsection{Humans' accuracy and reaction time depend on spatial frequency information}
\label{sec:results_psycghophsyics}
\begin{figure*}[!htb]
\begin{center}
\includegraphics[scale = 0.72]{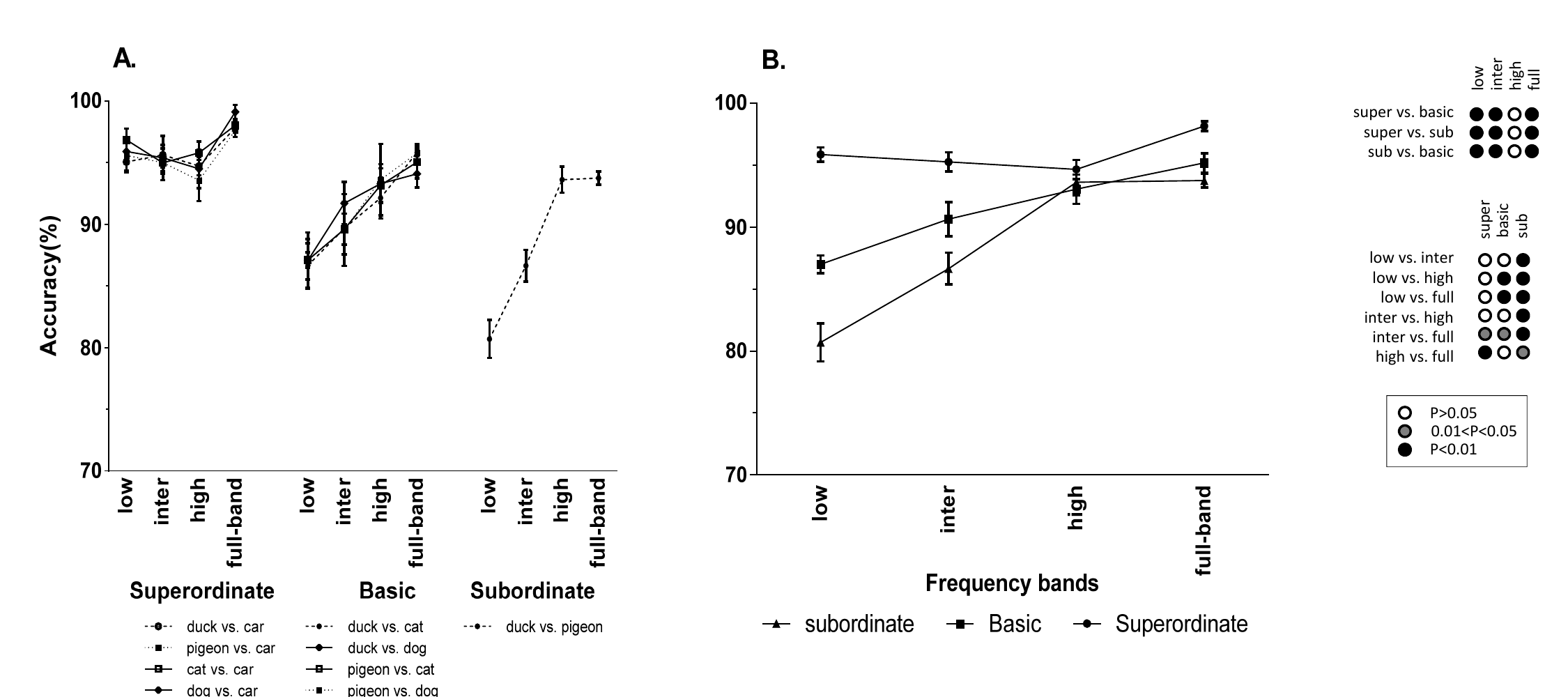}
\end{center}
\caption{
Subjects' mean accuracy for different categorization levels and spatial frequency bands.  A. Subjects' accuracy for the different tasks of each categorization level. B. Average accuracies over the different tasks of the superordinate (duck vs. car, pigeon vs. car, cat vs. car, and dog vs. car), basic (duck vs. cat, duck vs. dog, pigeon vs. cat, and pigeon vs. dog) and subordinate (duck vs. pigeon) levels. The p-value matrix presents the non-significant, significant and strongly significant values using white, gray, and black colors respectively. Error bars represent standard error of the mean (SEM).
}\label{fig:1}
\end{figure*}

The recognition accuracies of human subjects for the psychophysics experiments (see Section~\ref{sec:methods_psycghophsyics} for the details) with full-frequency band (i.e., original) images as well as the frequency-filtered images (i.e., LSF, ISF, and HSF bands) are shown in Figure~\ref{fig:1}. Figure~\ref{fig:1}A demonstrates the subjects mean accuracy for different tasks in each of the superordinate, basic, and subordinate level experiments. We performed three-factor ANOVA using spatial frequency, task, and categorization level as factors. This allows us to study the effect of each factor and their interactions on the categorization accuracy.

% The details of these experiments are introduced in Section~\ref{sec:methods_psycghophsyics}

As seen in the Superordinate column (left part) of Figure~\ref{fig:1}A, there is no significant difference between the accuracies in the four superordinate animal/non-animal tasks (p=0.542$>$0.05, F=0.720). Also, it can be seen that the categorization accuracies are very high in all the frequency bands. This means that the coarse information at LSF band is sufficient for the superordinate categorization level. The subjects' mean accuracy over each of the four (bird/non-bird) basic level tasks is presented in the Basic column (middle part) of Figure~\ref{fig:1}A. Here again, there is no significant difference among the four basic level tasks (p=0.958$>$0.05, F=0.103). However, the accuracy in basic level tasks dropped, with respect to the superordinate level. The maximum accuracy drop has occurred in the LSF band, while it is not changed much for the HSF band. The same effect is observed for the subordinate task, with higher accuracy drop in the LSF and ISF bands.

\begin{figure*}[!htb]
\begin{center}
\includegraphics[scale = 0.72]{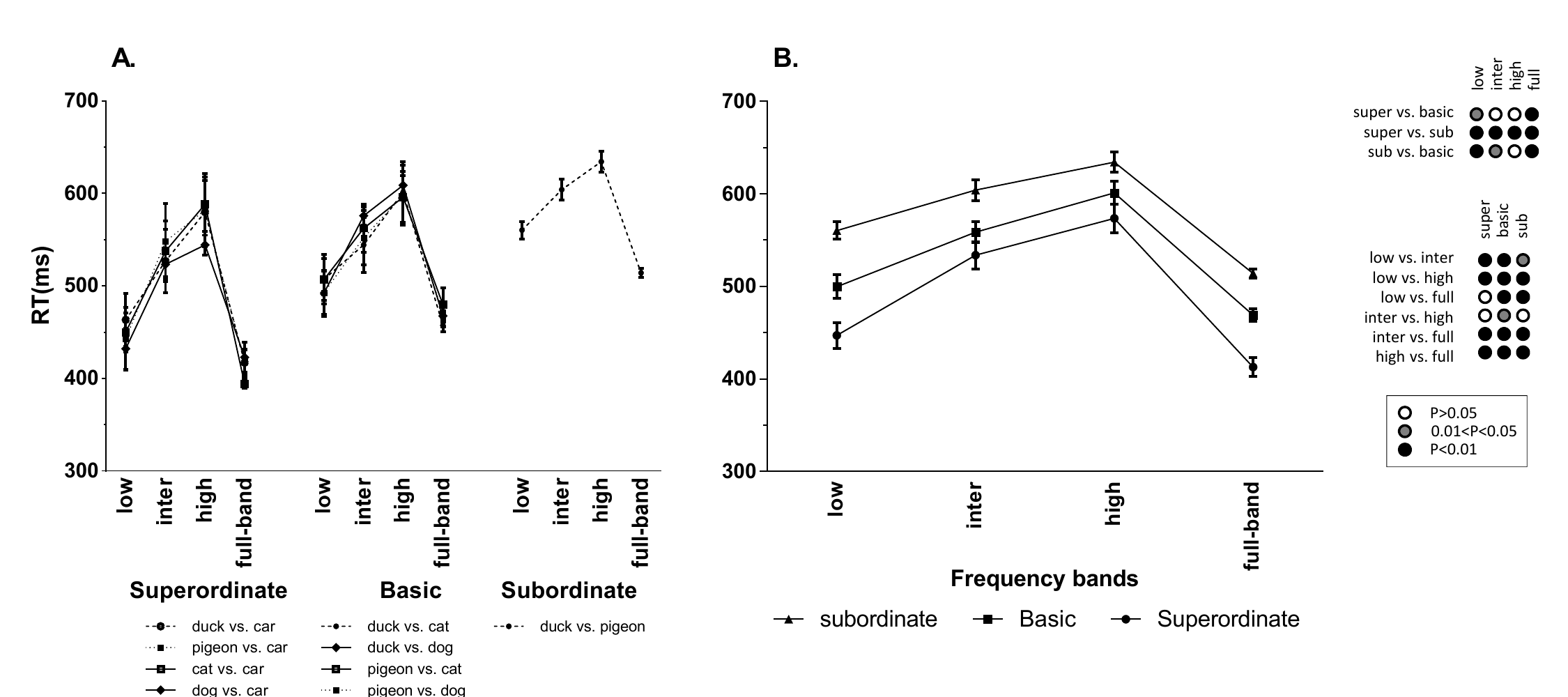}
\end{center}
\caption{
Subjects' mean RT for different categorization levels and spatial frequency bands.  A. Subjects' RT for the different tasks of each categorization level. B. Average RT over the different tasks of the superordinate (duck vs. car, pigeon vs. car, cat vs. car, and dog vs. car), basic (duck vs. cat, duck vs. dog, pigeon vs. cat, and pigeon vs. dog) and subordinate (duck vs. pigeon) levels. The p-value matrix presents the non-significant, significant and strongly significant values using white, gray, and black colors respectively. Error bars represent standard error of the mean (SEM).
}\label{fig:2}
\end{figure*}

Since there was no significant difference between the tasks corresponding to each categorization experiment, we also reported the average accuracy for each categorization level and each frequency band (see Figure~\ref{fig:1}B). For the superordinate level, by moving through the frequency bands from low to high, the accuracies decreased smoothly (this decrease is not statistically significant). In addition, the accuracy with the full-band images is significantly higher than the ISF and HSF bands, while the difference is not significant for LSF band (see p-value matrix in Figure~\ref{fig:1}). These together suggest that the LSF information is sufficient and necessary for superordinate level categorization. For the basic and subordinate levels, the LSF does not carry the required information by which the human subjects could precisely perform the categorization task. But, by shifting the frequency band toward the higher frequencies, the accuracy is constantly increasing. Compared to subordinate, the basic level has higher accuracy with lower frequency bands. Interestingly, for the HSF band, the accuracies corresponding to different categorization levels become closer to each other. Therefore, it can be said that the lower frequencies are suitable for higher categorization levels (e.g., superordinate), while performing low categorization levels (e.g., subordinate) require higher frequency information.

We also recorded the subjects RT during the experiments. Figures~\ref{fig:2}A presents the mean RT, independently for each psychophysics task. Like for the accuracies, we performed a three-factor ANOVA using the categorization level, frequency band, and task as independent factors. Again, there is no significant difference among the tasks in the superordinate (p-value=0.795$>$0.05; F=0.342) and basic (p-value=0.919$>$0.05; F=0.166) level experiments. Thus, in Figure~\ref{fig:2}B, we averaged the RTs of each categorization level and frequency band over the different tasks. For all categorization levels, the RT increases by moving from the low to the high ones. We also performed a two-factor ANOVA using frequency band and categorization level as factors. This indicates that the RTs of ISF experiments are significantly higher than those of LSF, and RTs of HSF experiments are significantly higher than those of ISF (see p-value matrix in Figure~\ref{fig:2}). These results are compatible with the previous findings indicating that lower frequencies are processed earlier than higher ones ~\cite{mace2005entry, kauffmann2015neural}.

On the other hand, there are several studies suggesting that the initial guess about the object is taken based on the LSF information which facilitates the recognition process in higher visual areas~\citep{fenske2006top, craddock2015early}. In agreement with these studies, the RTs of full-band experiments in all three categorization levels are significantly shorter than the HSF and ISF experiments, but close to the LSF ones (the differences are not statistically significant; see p-value matrix in Figure~\ref{fig:2}). Generally, the RTs of superordinate level are shorter than those of basic level, and RTs of basic level are shorter than those of subordinate level (see Figure~\ref{fig:2}B). In addition, the RT differences are longer in the LSF experiments. 

By considering both RTs (Figure~\ref{fig:2}B) and accuracies (Figure~\ref{fig:1}B), three main conclusions can be drawn. First, using just LSF information, humans could quickly and accurately accomplish the superordinate categorization. While, the RT is much longer with HSF information, despite the reasonable accuracy. These together suggests that the superordinate categorization is mainly done using the LSF information. Second, although categorization in basic and subordinate levels using the LSF information is completed faster, the accuracies are very low. Therefore, LSF information is not sufficient for the basic and subordinate levels. Intuitively, in the superordinate level, there is a high inter-category dissimilarity, and therefore, LSF information is sufficient for performing the task. For the subordinate tasks, with higher inter-category similarity, higher frequency information is required which carries more details about the object. Third, to complete the basic and subordinate level categorization, subjects needed HSF information. However, higher frequencies are processed later than lower ones. Hence, it can be concluded that superordinate level is the entry object categorization level, and then, categorization in the basic and subordinate levels are accomplished.  

\subsection{Computational models account for human behavior}
\label{sec:results_models}
\begin{figure*}[!htb]
\begin{center}
\includegraphics[scale = 0.65]{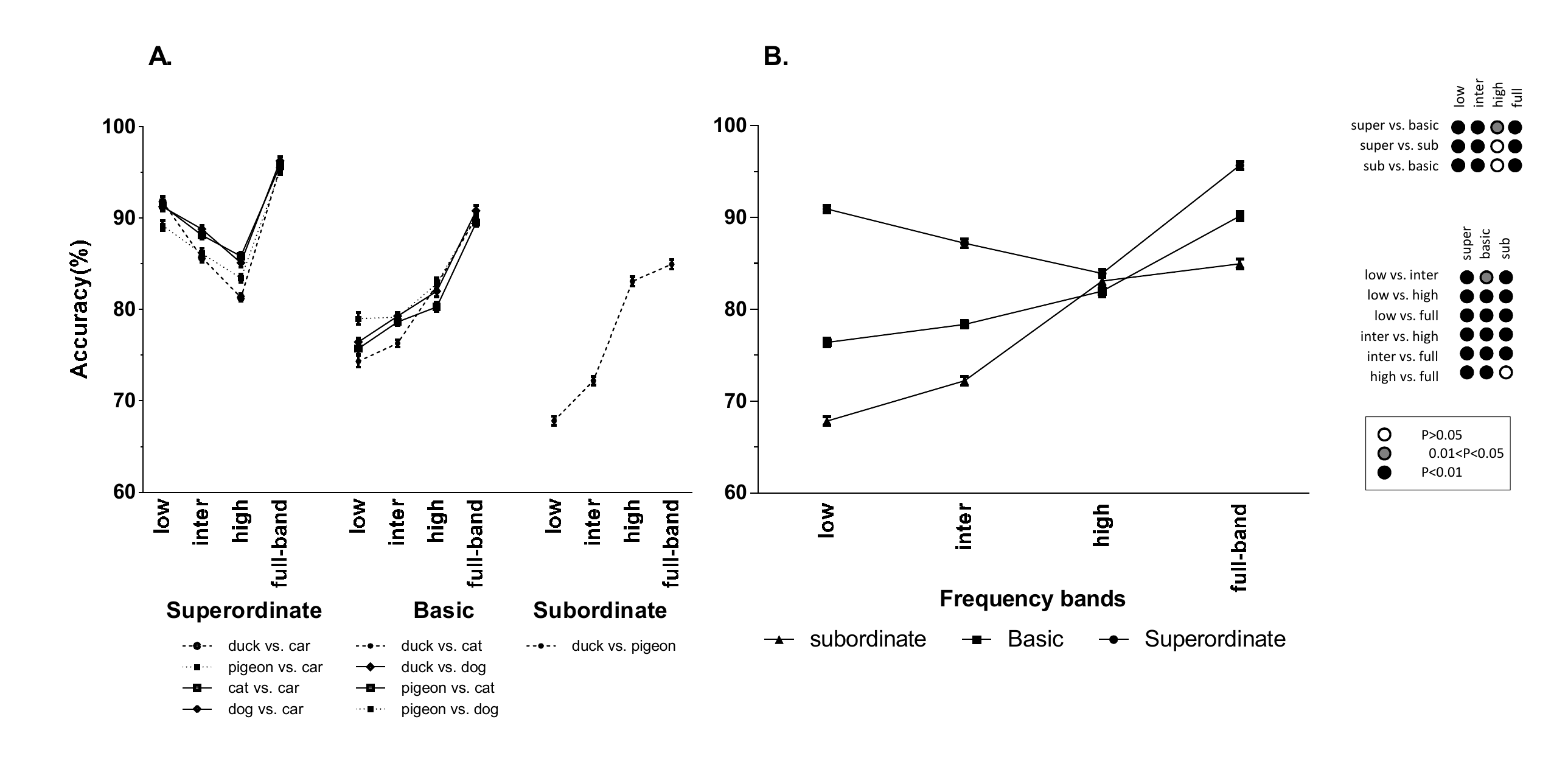}
\end{center}
\caption{
Mean accuracies of model~\rom{1}  (averaged over 10 independent runs) for different categorization levels and spatial frequency bands.  A. Model's accuracy for the different tasks of each categorization level. B. Average accuracies over the different tasks of the superordinate (duck vs. car, pigeon vs. car, cat vs. car, and dog vs. car), basic (duck vs. cat, duck vs. dog, pigeon vs. cat, and pigeon vs. dog) and subordinate (duck vs. pigeon) levels. The p-value matrix presents the non-significant, significant and strongly significant values using white, gray, and black colors respectively. Error bars represent standard error of the mean (SEM).
}\label{fig:3}
\end{figure*}

As mentioned in Section~\ref{sec:methods_models}, we employed two computational models to study whether the changes in human performance over the frequency bands are due to the changes in the information content in each band or due to the way the visual system processes different frequency information. Therefore, we assessed the models on the same categorization tasks as human subjects performed in the psychophysics experiments. The details of the models and the way they are trained and tested in each task are fully explained in Section~\ref{sec:methods_models}. Briefly, Gabor filters with different spatial frequencies are used in Model~\rom{1} to filter the input images into LSF, ISF, and HSF bands. Gabor filters with low (high) spatial frequencies act as low-pass (high-pass) filters and extract coarse (fine) information from the image. In Model~\rom{2}, images were directly filtered into the different frequency bands.

\begin{figure*}[!htb]
\begin{center}
\includegraphics[scale = 0.7]{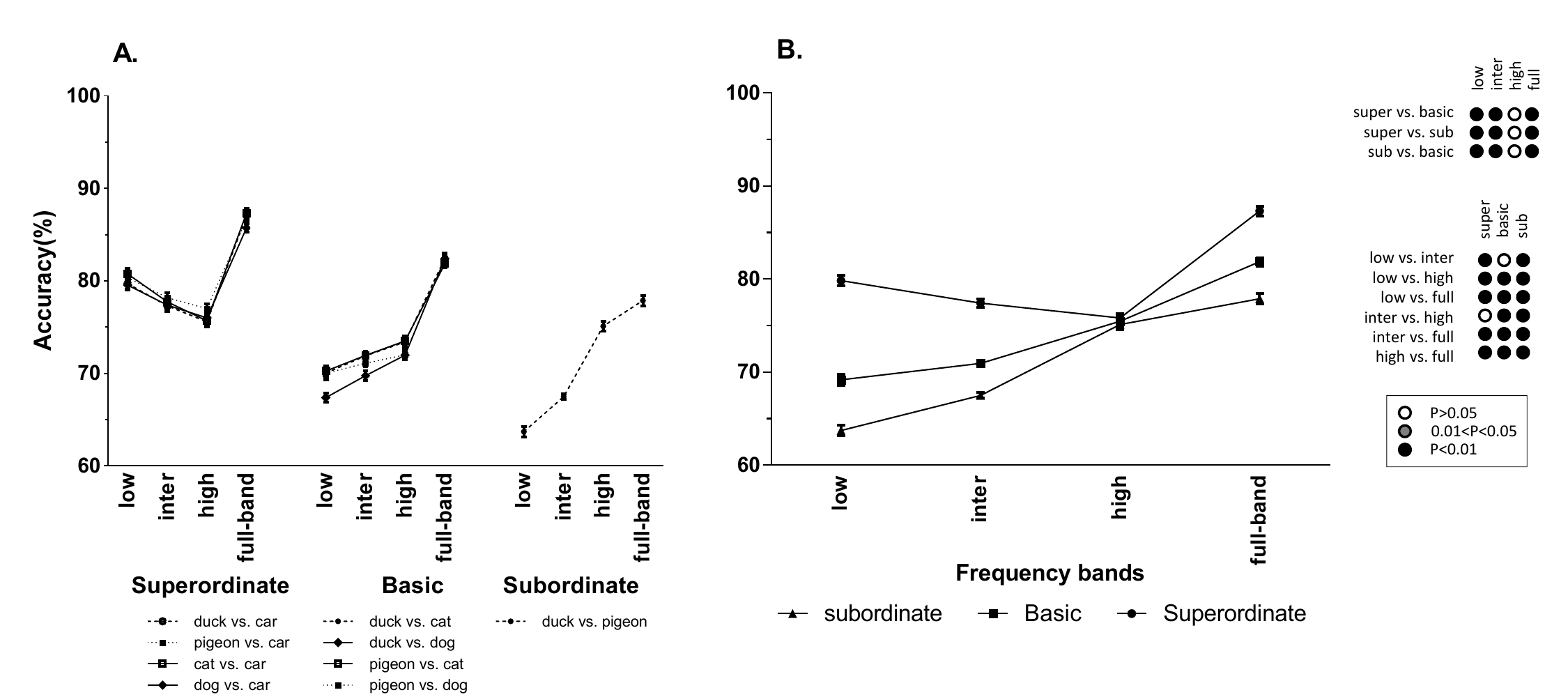}
\end{center}
\caption{
Mean accuracies of model~\rom{2}  (averaged over 10 independent runs) for different categorization levels and spatial frequency bands.  A. Model's accuracy for the different tasks of each categorization level. B. Average accuracies over the different tasks of the superordinate (duck vs. car, pigeon vs. car, cat vs. car, and dog vs. car), basic (duck vs. cat, duck vs. dog, pigeon vs. cat, and pigeon vs. dog) and subordinate (duck vs. pigeon) levels. The p-value matrix presents the non-significant, significant and strongly significant values using white, gray, and black colors respectively. Error bars represent standard error of the mean (SEM).
}\label{fig:4}
\end{figure*}

Figure~\ref{fig:3}A shows the accuracy of Model~\rom{1} over the different categorization tasks and levels, as well as frequency bands. Here again, we performed a three-factor ANOVA using categorization level, task, and frequency band as independent factors. Like for humans, there was no significant difference among the different tasks of each categorization level experiment (superordinate: p=0.091$>$0.05, F = 2.199; basic level: p=0.060$>$0.05, F = 2.525). Therefore, in Figure~\ref{fig:3}B, we averaged the accuracies across the categorization tasks. Interestingly, the overall trend in accuracies of the model~\rom{1} is very similar to those of humans (see Figure~\ref{fig:1}B): the accuracy of this model in superordinate level drops by moving from LSF band to the higher ones. Computationally, this means that lower frequencies contain more information about superordinate categories than the higher ones. While, in the basic and subordinate levels, higher frequency bands lead to higher accuracies. However, compared to the subordinate, the basic level has higher accuracies in LSF and ISF bands. Surprisingly, similarly to humans, the accuracies of all categorization levels become close to each other at the HSF band.

The accuracies of the model~\rom{2} are also presented in Figure~\ref{fig:4}, where Figure~\ref{fig:4} contains the accuracies on each task, and Figure~\ref{fig:4}B shows the accuracies averaged over the tasks. With respect to Model~\rom{1}, the accuracies of Model~\rom{2} have dropped, which is due to the elimination of the prepossessing stages (S1 and C1 layers) in this model. However, what matters is the trend of accuracies within and between the categorization levels. As seen, the results of model~\rom{2} are similar to those of model~\rom{1}. Again, LSF band lead to high accuracy in the superordinate level, while it results in the lowest accuracy for the other two categorization levels. Also, for accurate categorization in the basic and subordinate levels, HSF information is necessary.

Due to the consistency among the results of the two computational models with the human behavior, it can be said that the observed human accuracy pattern is mainly due to the information content in different frequency bands. In the LSF band, where the overall shape of the object is preserved and other details are removed, categorization in superordinate level can be done with high accuracy. This coarse information is not useful for the other two categorization levels (basic and subordinate), where the categories have more shape similarities. Therefore, the visual system needs more detailed information which lies in higher frequency bands. As stated before, lower spatial frequencies are processed faster than higher ones, and therefore, it is computationally difficult for the visual cortex to do superordinate categorization before
basic and subordinate levels.

Interestingly, similarly to humans, the accuracies of both models dropped in superordinate level, when moving from LSF band to the higher ones. Clearly, at the superordinate level, there is a huge variation among the objects in each category. Therefore, lower frequencies, which maintain the overall shape of the objects, contain the required information for superordinate level categorization. However, in the basic and subordinate levels, the higher frequency bands are more informative. Compared to the subordinate level, the basic level has higher accuracies in LSF and ISF bands. Surprisingly, similarly to humans, the accuracies of all the categorization levels meet each other at the HSF band.

\begin{figure*}[!htb]
\begin{center}
\includegraphics[scale = 0.5]{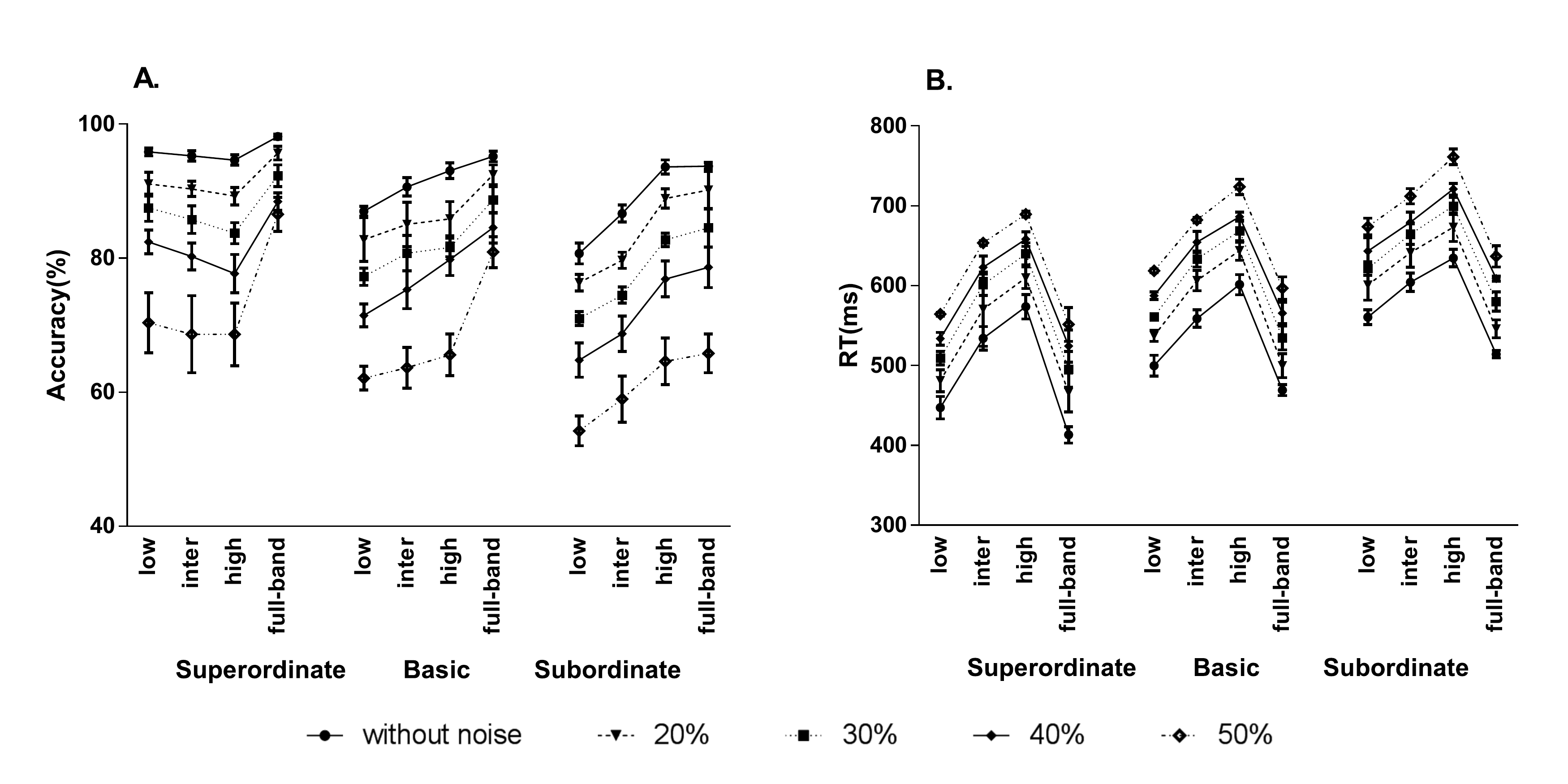}
\end{center}
\caption{
The effect of adding different amounts of noise to the image stimuli (20\%, 30\%, 40\%, and 50\%) on humans' accuracy (part A) and RT (part B).  Error bars represent standard error of the mean (SEM).
}\label{fig:5}
\end{figure*}

\subsection{Results are robust to noise}
\label{sec:results_noise}
We repeated all the behavioral (see Section~\ref{sec:methods_psycghophsyics}) and computational (see Section~\ref{sec:methods_models}) experiments with noisy images described in Section~\ref{sec:dataset}. Adding noise to the input image will increase the errors, and therefore, it will avoid any potential ceiling effects in the performances. Also, it allows us to check whether the obtained results are still valid under more difficult visual conditions.

Figure~\ref{fig:5}A and \ref{fig:5}B show the human subjects' categorization accuracy and RT for the full-band and frequency-filtered images with different amounts of noise, respectively. Note that, for each categorization level experiment, we averaged the accuracies and RTs corresponding to each task. For instance, for the subordinate level experiment, there was four animal/non-animal tasks (see section~\ref{sec:dataset}). 

As shown in Figure~\ref{fig:5}A, when increasing the noise level, the accuracy dropped. But, the overall trend of accuracies over the frequency bands and categorization levels remains the same. For the superordinate categorization, LSF band has higher accuracy than intermediate and high bands, even for 50\% noise level. For the basic and subordinate levels, the accuracy increases moving from LSF to ISF and HSF bands. To statistically investigate the effect of noise on categorization accuracies, again we performed a three-factor ANOVA using noise, categorization level, and frequency band as independent factors. Our analysis indicates that there is no significant interaction between the categorization and noise levels, meaning that adding noise has no effect on the accuracy trend over the categorization levels. However, the noise level has a significant effect on the accuracy (p=0.0001$<$0.05, F=243.759). All the other effects and interactions were also significant (frequency:p=0.0001$<$0.05, F=74.261; categorization level:p=0.0001$<$0.05, F=99.628; frequency$\times$categorization level: p=0.0001$<$0.05, F=10.636; frequency$\times$ noise:p=0.038$<$0.05, F=1.858).

\begin{figure*}[!htb]
\begin{center}
\includegraphics[scale = 0.5]{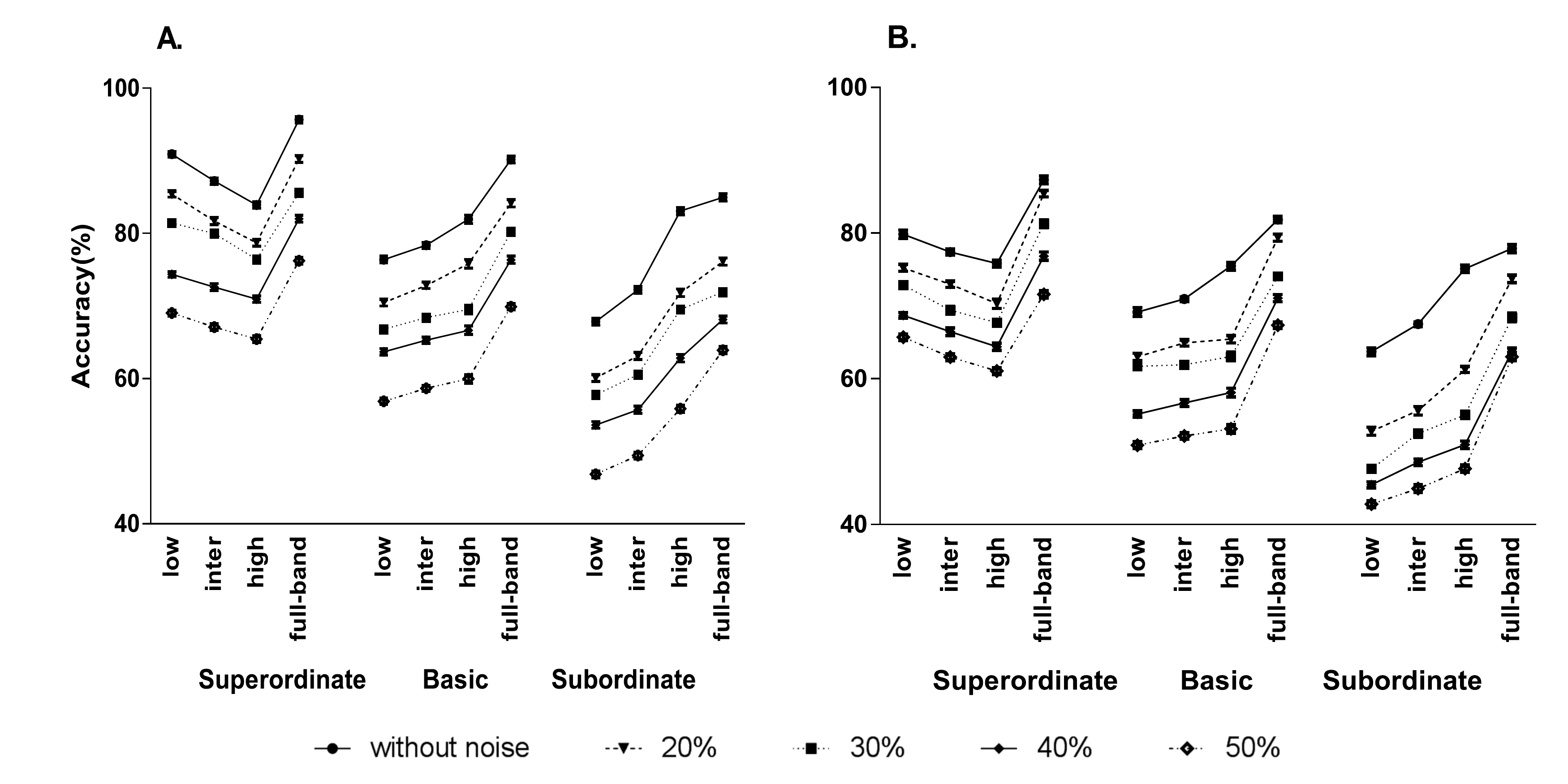}
\end{center}
\caption{
The effect of adding different amounts of noise to the image stimuli (20\%, 30\%, 40\%, and 50\%) on the accuracy of model~\rom{1} (part A) and model~\rom{2} (part B).  Error bars represent standard error of the mean (SEM).
}\label{fig:6}
\end{figure*}

Adding noise also increases the RT (see Figure~\ref{fig:5}B). Similarly to the noise-free experiments, for all categorization levels, the RT increases by moving from  LSF to ISF and HSF bands. Here again, subordinate (basic) has longer RTs than basic (superordinate) level. Interestingly, for all categorization levels, as the amount of noise increases, the pattern of RTs is maintained but shifted toward longer times. We performed a three-factor ANOVA to study the impacts of noise, categorization level, and frequency band on RTs. Similarly to the accuracies, noise had a significant effect on RTs (p=0.0001$<$0.05, F=156.275), and there was no significant interaction between the categorization level and noise. All the other effects and interactions were also significant (frequency:p=0.0001$<$0.05, F=327.685; categorization level:p=0.0001$<$0.05, F=202.089; frequency$\times$categorization level:p=0.0001$<$0.05, F=4.144).

Also, we evaluated the two computational models on the noisy images. This allowed to study, from a computational point of view, how adding noise affects the information at each frequency band, i.e., accuracy in each categorization level. The categorization accuracies of Model~\rom{1} and Model~\rom{2} for different levels of noise are shown in  Figure~\ref{fig:6}A and~\ref{fig:6}B, respectively. Similarly to the humans, by increasing the amount of noise, the accuracies dropped, while the trend of accuracies over the frequency bands was maintained. In addition, for all noise levels, LSF band leads to higher accuracies in superordinate level than ISF and HSF bands, while higher frequencies are suitable for basic and subordinate levels.
 %These together confirms that noise addition  

\section{Discussion}
The aim of our study was to investigate the effect of spatial frequencies on object categorization in different levels (i.e., superordinate, basic and subordinate levels). We constructed an object dataset containing images of cars and four animals (ducks, pigeons, cats, and dogs). Images were also filtered into LSF, ISF, and HSF bands. We performed several rapid-presentation psychophysics experiments at different categorization levels, and recorded the subjects' accuracy and RT. The same categorization experiments were also performed by two computational models .
%For each level, we had different categorization tasks. For instance, we had four animal/non-animal tasks (one animal v.s. car), in superordinate level.

Although, the relation between the global (resp. local) visual processing and LSF (resp. HSF) information has been debated~\citep{morrison2001usage,loftus2004different,goffaux2006faces,goffaux2005respective,boutet2003configural}, coarse information is obviously excluded from the HSF-filtered images, whereas the fine details such as sharp edges and textures are absent in LSF-filtered images. Results of our psychophysics experiments reveal that
%subordinate categorization mainly relies on high frequency information (i.e.,  local details), while for the general categorization, low frequency information (i.e., global features) is sufficient. In other words, our results show that, 
using just low spatial frequencies, humans could accurately perform superordinate level categorization. While, for the basic and subordinate levels, higher frequency information is required and humans could not reach high precisions using just LSF information. In addition, the accuracy at basic level was greater than the subordinate level for LSF and ISF bands. These together indicate that basic level is not fully dependent on HSF or LSF bands. Also, at HSF band, the human accuracy in all categorization levels is almost equal.
%While the RT constantly increases from superordinate to  basic level and from basic to subordinate level.
 This suggests that HSF bands carry the same amount of information useful for each categorization level.

However, these results are contrary to the findings of~\cite{collin2005subordinate}, where they suggested that the highest accuracy with LSF information is reached at the basic level, while superordinate and subordinate levels rely on higher frequencies. This contradiction could be due to the employed speeded category verification task which included more analysis beyond the object detection. Actually, they presented a category name followed by an object image, and subjects had to report if they matched or not. Reading and language cortical areas might be activated during the visual process which can give advantage to the basic level, over the superordinate level. Therefore, this type of experiment has been criticized because of involving the semantic processing of the brain~\citep{wu2014120,mace2009time, rosch1976basic}.

Our result are compatible with the theory of coarse-to-fine temporal processing in visual system~\citep{ navon1977forest, schyns1994blobs, hughes1996global, mace2005entry}, where lower frequencies are processed earlier than higher ones, independently of the categorization level. In addition, the subjects' RT for superordinate  (resp. basic)  level was shorter than  the basic (resp. subordinate) level. These suggest that the superordinate level is the entry  categorization level, and subordinate level is the latest one. However, these results contradicts with the findings of some earlier studies~\citep{large2004electrophysiological, gauthier1997levels, murphy1985category, murphy1989categorizing, jolicoeur1984pictures} suggesting the temporal advantage for the basic level categorization. Particularly, \cite{collin2005subordinate} suggested that, whatever the spatial frequency band, the basic level is completed earlier than superordinate and subordinate levels. Again, this could be due to the employed speeded category verification paradigm and long stimulus presentation time in their experiments.

From a computational point of view, our results could be due to the information content at each frequency band, or the underlying neural processes involving object categorization at each level. We used two computational models to do the same categorization experiments as humans did. These models employ different frequency filtering mechanisms; one uses Gabor filters with low to high spatial frequencies, and the other one uses directly filtered images. Both models reached similar accuracy pattern to those of humans over the different categorization levels and frequency bands. Therefore, computationally, neither basic nor subordinate level can be the entry level, due to the lack of required visual information in the LSF band.

%\bibliographystyle{frontiersinSCNS_ENG_HUMS} % for Science, Engineering and Humanities and Social Sciences articles, for Humanities and Social Sciences articles please include page numbers in the in-text citations
%%\bibliographystyle{frontiersinHLTH&FPHY} % for Health, Physics and Mathematics articles
%\bibliography{test}

\begin{thebibliography}{35}
\providecommand{\natexlab}[1]{#1}
\expandafter\ifx\csname urlstyle\endcsname\relax
  \providecommand{\doi}[1]{doi:\discretionary{}{}{}#1}\else
  \providecommand{\doi}{doi:\discretionary{}{}{}\begingroup
  \urlstyle{rm}\Url}\fi
\providecommand{\selectlanguage}[1]{\relax}
\providecommand{\bibAnnoteFile}[1]{%
  \IfFileExists{#1}{\begin{quotation}\noindent\textsc{Key:} #1\\
  \textsc{Annotation:}\ \input{#1}\end{quotation}}{}}
\providecommand{\bibAnnote}[2]{%
  \begin{quotation}\noindent\textsc{Key:} #1\\
  \textsc{Annotation:}\ #2\end{quotation}}

\bibitem[{Ales et~al.(2012)Ales, Farzin, Rossion, and
  Norcia}]{ales2012objective}
Ales, J.~M., Farzin, F., Rossion, B., and Norcia, A.~M. (2012).
\newblock An objective method for measuring face detection thresholds using the
  sweep steady-state visual evoked response.
\newblock \emph{Journal of vision} 12, 18--18
\input{ales2012objective}

\bibitem[{Bar et~al.(2006)Bar, Kassam, Ghuman, Boshyan, Schmid, Dale
  et~al.}]{bar2006top}
Bar, M., Kassam, K.~S., Ghuman, A.~S., Boshyan, J., Schmid, A.~M., Dale, A.~M.,
  et~al. (2006).
\newblock Top-down facilitation of visual recognition.
\newblock \emph{Proceedings of the National Academy of Sciences of the United
  States of America} 103, 449--454
\input{bar2006top}

\bibitem[{Boutet et~al.(2003)Boutet, Collin, and
  Faubert}]{boutet2003configural}
Boutet, I., Collin, C., and Faubert, J. (2003).
\newblock Configural face encoding and spatial frequency information.
\newblock \emph{Attention, Perception, \& Psychophysics} 65, 1078--1093
\input{boutet2003configural}

\bibitem[{Bowers and Jones(2008)}]{bowers2008detecting}
Bowers, J.~S. and Jones, K.~W. (2008).
\newblock Detecting objects is easier than categorizing them.
\newblock \emph{The Quarterly journal of experimental psychology} 61, 552--557
\input{bowers2008detecting}

\bibitem[{Brainard(1997)}]{brainard1997psychophysics}
Brainard, D.~H. (1997).
\newblock The psychophysics toolbox.
\newblock \emph{Spatial vision} 10, 433--436
\input{brainard1997psychophysics}

\bibitem[{Collin and Mcmullen(2005)}]{collin2005subordinate}
Collin, C.~A. and Mcmullen, P.~A. (2005).
\newblock Subordinate-level categorization relies on high spatial frequencies
  to a greater degree than basic-level categorization.
\newblock \emph{Attention, Perception, \& Psychophysics} 67, 354--364
\input{collin2005subordinate}

\bibitem[{Craddock et~al.(2015)Craddock, Martinovic, and
  M{\"u}ller}]{craddock2015early}
Craddock, M., Martinovic, J., and M{\"u}ller, M.~M. (2015).
\newblock Early and late effects of objecthood and spatial frequency on
  event-related potentials and gamma band activity.
\newblock \emph{BMC neuroscience} 16, 6
\input{craddock2015early}

\bibitem[{Dehaqani et~al.(2016)Dehaqani, Vahabie, Kiani, Ahmadabadi, Araabi,
  and Esteky}]{dehaqani2016temporal}
Dehaqani, M.-R.~A., Vahabie, A.-H., Kiani, R., Ahmadabadi, M.~N., Araabi,
  B.~N., and Esteky, H. (2016).
\newblock Temporal dynamics of visual category representation in the macaque
  inferior temporal cortex.
\newblock \emph{Journal of neurophysiology} , jn--00018
\input{dehaqani2016temporal}

\bibitem[{Fenske et~al.(2006)Fenske, Aminoff, Gronau, and Bar}]{fenske2006top}
Fenske, M.~J., Aminoff, E., Gronau, N., and Bar, M. (2006).
\newblock Top-down facilitation of visual object recognition: object-based and
  context-based contributions.
\newblock \emph{Progress in brain research} 155, 3--21
\input{fenske2006top}

\bibitem[{Gauthier et~al.(1997)Gauthier, Anderson, Tarr, Skudlarski, and
  Gore}]{gauthier1997levels}
Gauthier, I., Anderson, A.~W., Tarr, M.~J., Skudlarski, P., and Gore, J.~C.
  (1997).
\newblock Levels of categorization in visual recognition studied using
  functional magnetic resonance imaging.
\newblock \emph{Current Biology} 7, 645--651
\input{gauthier1997levels}

\bibitem[{Goffaux et~al.(2005)Goffaux, Hault, Michel, Vuong, and
  Rossion}]{goffaux2005respective}
Goffaux, V., Hault, B., Michel, C., Vuong, Q.~C., and Rossion, B. (2005).
\newblock The respective role of low and high spatial frequencies in supporting
  configural and featural processing of faces.
\newblock \emph{Perception} 34, 77--86
\input{goffaux2005respective}

\bibitem[{Goffaux and Rossion(2006)}]{goffaux2006faces}
Goffaux, V. and Rossion, B. (2006).
\newblock Faces are" spatial"--holistic face perception is supported by low
  spatial frequencies.
\newblock \emph{Journal of Experimental Psychology: Human Perception and
  Performance} 32, 1023
\input{goffaux2006faces}

\bibitem[{Hughes et~al.(1996)Hughes, Nozawa, and Kitterle}]{hughes1996global}
Hughes, H.~C., Nozawa, G., and Kitterle, F. (1996).
\newblock Global precedence, spatial frequency channels, and the statistics of
  natural images.
\newblock \emph{Journal of cognitive neuroscience} 8, 197--230
\input{hughes1996global}

\bibitem[{Jolicoeur et~al.(1984)Jolicoeur, Gluck, and
  Kosslyn}]{jolicoeur1984pictures}
Jolicoeur, P., Gluck, M.~A., and Kosslyn, S.~M. (1984).
\newblock Pictures and names: Making the connection.
\newblock \emph{Cognitive psychology} 16, 243--275
\input{jolicoeur1984pictures}

\bibitem[{Kauffmann et~al.(2015)Kauffmann, Bourgin, Guyader, and
  Peyrin}]{kauffmann2015neural}
Kauffmann, L., Bourgin, J., Guyader, N., and Peyrin, C. (2015).
\newblock The neural bases of the semantic interference of spatial
  frequency-based information in scenes.
\newblock \emph{Journal of cognitive neuroscience}
\input{kauffmann2015neural}

\bibitem[{Kauffmann et~al.(2014)Kauffmann, Ramano{\"e}l, and
  Peyrin}]{kauffmann2014neural}
Kauffmann, L., Ramano{\"e}l, S., and Peyrin, C. (2014).
\newblock The neural bases of spatial frequency processing during scene
  perception.
\newblock \emph{Frontiers in Integrative Neuroscience} 8
\input{kauffmann2014neural}

\bibitem[{Large et~al.(2004)Large, Kiss, and
  McMullen}]{large2004electrophysiological}
Large, M.-E., Kiss, I., and McMullen, P.~A. (2004).
\newblock Electrophysiological correlates of object categorization: Back to
  basics.
\newblock \emph{Cognitive Brain Research} 20, 415--426
\input{large2004electrophysiological}

\bibitem[{Loftus and Harley(2004)}]{loftus2004different}
Loftus, G.~R. and Harley, E.~M. (2004).
\newblock How different spatial-frequency components contribute to visual
  information acquisition.
\newblock \emph{Journal of Experimental Psychology: Human Perception and
  Performance} 30, 104
\input{loftus2004different}

\bibitem[{Mac{\'e} et~al.(2005)Mac{\'e}, Joubert, and
  Fabre~Thorpe}]{mace2005entry}
Mac{\'e}, M., Joubert, O., and Fabre~Thorpe, M. (2005).
\newblock Entry level at the superordinate level in visual categorization.
\newblock In \emph{9th International conference on cognitive and neural
  systems}. vol.~52
\input{mace2005entry}

\bibitem[{Mac{\'e} et~al.(2009)Mac{\'e}, Joubert, Nespoulous, and
  Fabre-Thorpe}]{mace2009time}
Mac{\'e}, M. J.-M., Joubert, O.~R., Nespoulous, J.-L., and Fabre-Thorpe, M.
  (2009).
\newblock The time-course of visual categorizations: you spot the animal faster
  than the bird.
\newblock \emph{PloS one} 4, e5927
\input{mace2009time}

\bibitem[{Mack and Palmeri(2015)}]{mack2015dynamics}
Mack, M.~L. and Palmeri, T.~J. (2015).
\newblock The dynamics of categorization: Unraveling rapid categorization.
\newblock \emph{Journal of Experimental Psychology: General} 144, 551
\input{mack2015dynamics}

\bibitem[{Morrison and Schyns(2001)}]{morrison2001usage}
Morrison, D.~J. and Schyns, P.~G. (2001).
\newblock Usage of spatial scales for the categorization of faces, objects, and
  scenes.
\newblock \emph{Psychonomic Bulletin \& Review} 8, 454--469
\input{morrison2001usage}

\bibitem[{Murphy and Brownell(1985)}]{murphy1985category}
Murphy, G.~L. and Brownell, H.~H. (1985).
\newblock Category differentiation in object recognition: typicality
  constraints on the basic category advantage.
\newblock \emph{Journal of Experimental Psychology: Learning, Memory, and
  Cognition} 11, 70
\input{murphy1985category}

\bibitem[{Murphy and Wisniewski(1989)}]{murphy1989categorizing}
Murphy, G.~L. and Wisniewski, E.~J. (1989).
\newblock Categorizing objects in isolation and in scenes: What a superordinate
  is good for.
\newblock \emph{Journal of Experimental Psychology: Learning, Memory, and
  Cognition} 15, 572--586
\input{murphy1989categorizing}

\bibitem[{Navon(1977)}]{navon1977forest}
Navon, D. (1977).
\newblock Forest before trees: The precedence of global features in visual
  perception.
\newblock \emph{Cognitive psychology} 9, 353--383
\input{navon1977forest}

\bibitem[{Poncet and Fabre-Thorpe(2014)}]{poncet2014stimulus}
Poncet, M. and Fabre-Thorpe, M. (2014).
\newblock Stimulus duration and diversity do not reverse the advantage for
  superordinate-level representations: the animal is seen before the bird.
\newblock \emph{European Journal of Neuroscience} 39, 1508--1516
\input{poncet2014stimulus}

\bibitem[{Pra{\ss} et~al.(2013)Pra{\ss}, Grimsen, K{\"o}nig, and
  Fahle}]{prass2013ultra}
Pra{\ss}, M., Grimsen, C., K{\"o}nig, M., and Fahle, M. (2013).
\newblock Ultra rapid object categorization: effects of level, animacy and
  context.
\newblock \emph{PLoS One} 8, e68051
\input{prass2013ultra}

\bibitem[{Rogers and Patterson(2007)}]{rogers2007object}
Rogers, T.~T. and Patterson, K. (2007).
\newblock Object categorization: reversals and explanations of the basic-level
  advantage.
\newblock \emph{Journal of Experimental Psychology: General} 136, 451
\input{rogers2007object}

\bibitem[{Rosch et~al.(1976)Rosch, Mervis, Gray, Johnson, and
  Boyes-Braem}]{rosch1976basic}
Rosch, E., Mervis, C.~B., Gray, W.~D., Johnson, D.~M., and Boyes-Braem, P.
  (1976).
\newblock Basic objects in natural categories.
\newblock \emph{Cognitive psychology} 8, 382--439
\input{rosch1976basic}

\bibitem[{Russakovsky et~al.(2015)Russakovsky, Deng, Su, Krause, Satheesh, Ma
  et~al.}]{ILSVRC15}
Russakovsky, O., Deng, J., Su, H., Krause, J., Satheesh, S., Ma, S., et~al.
  (2015).
\newblock {ImageNet Large Scale Visual Recognition Challenge}.
\newblock \emph{International Journal of Computer Vision (IJCV)} 115, 211--252.
\newblock \doi{10.1007/s11263-015-0816-y}
\input{ILSVRC15}

\bibitem[{Schyns and Oliva(1994)}]{schyns1994blobs}
Schyns, P.~G. and Oliva, A. (1994).
\newblock From blobs to boundary edges: Evidence for time-and
  spatial-scale-dependent scene recognition.
\newblock \emph{Psychological science} 5, 195--200
\input{schyns1994blobs}

\bibitem[{Serre et~al.(2007)Serre, Wolf, Bileschi, Riesenhuber, and
  Poggio}]{serre2007robust}
Serre, T., Wolf, L., Bileschi, S., Riesenhuber, M., and Poggio, T. (2007).
\newblock Robust object recognition with cortex-like mechanisms.
\newblock \emph{IEEE transactions on pattern analysis and machine intelligence}
  29
\input{serre2007robust}

\bibitem[{Tanaka and Taylor(1991)}]{tanaka1991object}
Tanaka, J.~W. and Taylor, M. (1991).
\newblock Object categories and expertise: Is the basic level in the eye of the
  beholder?
\newblock \emph{Cognitive psychology} 23, 457--482
\input{tanaka1991object}

\bibitem[{Vanmarcke et~al.(2016)Vanmarcke, Calders, and
  Wagemans}]{vanmarcke2016time}
Vanmarcke, S., Calders, F., and Wagemans, J. (2016).
\newblock The time-course of ultrarapid categorization: The influence of scene
  congruency and top-down processing.
\newblock \emph{i-Perception} 7, 2041669516673384
\input{vanmarcke2016time}

\bibitem[{Wu et~al.(2014)Wu, Crouzet, Thorpe, and Fabre-Thorpe}]{wu2014120}
Wu, C.-T., Crouzet, S.~M., Thorpe, S.~J., and Fabre-Thorpe, M. (2014).
\newblock At 120 msec you can spot the animal but you don't yet know it's a
  dog.
\newblock \emph{Journal of Cognitive Neuroscience}
\input{wu2014120}

\end{thebibliography}

\end{document}